%% file: main.tex
\documentclass[%
reprint,
superscriptaddress,
 amsmath,amssymb,
 aps,
pre,
floatfix,
longbibliography
]{revtex4-1}
\usepackage{graphicx}
\usepackage[caption=false]{subfig}
\usepackage{dcolumn}
\usepackage{bm}
\usepackage{tabularx}

\usepackage{color}

\begin{document}



\title{Double origin of stochastic granular tribocharging}

\author{Jan Haeberle}
\affiliation{Institut f\"ur Materialphysik im Weltraum, Deutsches Zentrum f\"ur Luft- und Raumfahrt, 51170 K\"oln, Germany}

\author{Andr\'{e} Schella}
\affiliation{Max-Planck-Institute for Dynamics and Self-Organization G\"ottingen, 37077 G\"ottingen, Germany}

\author{Matthias Sperl}
\affiliation{Institut f\"ur Materialphysik im Weltraum, Deutsches Zentrum f\"ur Luft- und Raumfahrt, 51170 K\"oln, Germany}
\affiliation{Institut f\"ur Theoretische Physik, Universit\"at zu K\"oln, 50937 K\"oln, Germany}

\author{Matthias Schr\"oter}
\affiliation{Max-Planck-Institute for Dynamics and Self-Organization G\"ottingen, 37077 G\"ottingen, Germany}
\affiliation{Institute for Multiscale Simulation, Friedrich-Alexander-Universit\"at Erlangen-N\"urnberg (FAU), 91052 Erlangen, Germany}

\author{Philip Born}
\email{philip.born@dlr.de}
\affiliation{Institut f\"ur Materialphysik im Weltraum, Deutsches Zentrum f\"ur Luft- und Raumfahrt, 51170 K\"oln, Germany}

\date{\today}

\begin{abstract}
The mechanisms underlying triboelectric charging have a stochastic nature. We investigate how this randomness affects the distributions of charges generated on granular particles during either a single or many collisions. The charge distributions we find in our experiments are more heavy-tailed than normal distributions with an exponential decay of the probability, they are asymmetric, and exhibit charges of both signs. Moreover, we find a linear correlation between the width and mean of these distributions. We rationalize these findings with a model for triboelectric charging which combines stochastic charge separation during contact and stochastic charge recombination after separation of the surfaces. Our results further imply that subsequent charging events are not statistically independent.
 
\end{abstract}

\maketitle



\section{\label{sec:intro}Introduction}

Forming and breaking of contacts among solid bodies is intrinsically connected to generation of electrostatic charge \cite{shaw_experiments_1917,loeb_basic_1945,harper_electrification_1961}. This contact- or triboelectric charging has many spectacular manifestations in granular media, among which are flashes in volcano plumes \cite{cimarelli_multiparametric_2016,cimarelli_experimental_2014}, lightnings in sand storms \cite{stow_dust_1969} and self-ignition of dust explosions \cite{glor_hazards_1985, gibson_incendivity_1965}. Triboelectric charging in granular media has also found technical implementations as in photocopying \cite{pai_physics_1993}, electrostatic powder coating \cite{bailey_science_1998} or electrostatic dust removal \cite{mizuno_electrostatic_2000,kawamoto_mitigation_2011,calle_active_2011}. Despite the widespread occurrence of triboelectric charging,  no generally accepted theoretical framework has been developed for all these effects.  Experimental evidence suggests that the charge separation occurs by a wealth of mechanisms, out of which individual mechanisms may prevail under certain circumstances \cite{waitukaitis_size-dependent_2014,ducati_charge_2010,shinbrot_spontaneous_2008}, and quantitative predictions seem to be out of reach \cite{baytekin_mosaic_2011, lacks_unpredictability_2012}.

A joint feature of the mechanisms proposed to build up and dissipate static charging is their stochastic nature (see discussion in sec.~\ref{sec:theoryI}). Here we focus on this stochastic nature of triboelectric charging of dielectric particles. Knowing the characteristics of the probability distribution of the charges is of interest in various situations. The probability of igniting a spark depends on the probability of accumulating an extreme charge in a contact between particles, while the efficiency of coating processes and dust removal may be better derived from the average charge of the particles. Correct modeling of particle interaction depends on the whole range of accessible charges.

In this work, we study the probability distribution of triboelectric charging in both single collision experiments and for many subsequent collisions (see sec.~\ref{sec:exp}). We find asymmetric, exponential-tailed distributions which range from positive to negative charges as a common feature in all our experiments. Additionally, the experimental results imply that subsequent charging events are not statistically independent and that the mean and the width of the charge distributions are linearly correlated. 

These results motivated a search for a common underlying stochastic mechanisms. In section \ref{sec:theoryII} we suggest a model for the probability distribution of triboelectric charges based on the two stochastic triboelectric processes, charge separation and charge dissipation. The model reproduces the general features of 
our measured charge distributions which suggests that charging and discharging are equally relevant for understanding triboelectricity. 


\section{\label{sec:theoryI} Mechanisms of triboelectricity} 

The charge build-up upon breakage of contact of solid bodies is often described in a first approximation as a material property. This may be motivated by the well-understood contact charging of metals, where charge build-up can be predicted by the work functions of the materials \cite{lowell_contact_1980,lowell_triboelectrification_1986}. Based on this material-focused view,  triboelectric series, which rank the affinity of a material to charge positively or negatively after a contact \cite{shaw_experiments_1917}, have been developed for a number of materials including insulators.

However, inconsistencies among different reported triboelectric series can be found repeatedly \cite{diaz_semi-quantitative_2004,park_triboelectric_2008,mccarty_electrostatic_2008}. This problem motivated the search for further parameters relevant to charge separation. Experiments identified, among others, hydrophobicity \cite{schella_influence_2017,lee_collisional_2018}, humidity \cite{nemeth_polymer_2003,howell_dynamics_2001, xie_effect_2016}, temperature \cite{poppe_further_2005}, strain \cite{sow_strain-induced_2012}, particle size \cite{waitukaitis_size-dependent_2014,lacks_nonequilibrium_2008}, impact velocity and angle \cite{matsuyama_charge_1995, matsusyama_impact_2006,xie_effect_2016} and particle shape and contact mode \cite{ireland_triboelectrification_2010,ireland_triboelectrification_2010-1,ireland_dynamic_2012} as relevant parameters influencing triboelectric charging.

Four mechanisms are presently discussed to fundamentally cause charge separation in contacts of insulators. One mechanism is the exchange of electrons trapped in localized states within the band gap of the insulators \cite{lowell_contact_1980,lowell_triboelectrification_1986,wood_weak_1972}. Electrons can relax from such excited trapped states near the surface of one body into states in the valence band of another body in contact, such that a net charge remains after separation of the bodies. The energy levels and the frequency of these trapped states are randomly distributed \cite{wood_weak_1972}, and an additional probability for  a relaxation process has to be taken into account. It should be pointed out that recent work has questioned the generality of the trapped states model \cite{waitukaitis_size-dependent_2014}.

Another mechanism that separates charges is the exchange of mobile ions and ion exchange through a medium. If mobile ions are present on the surfaces in contact, the concentrations will equilibrate by thermal motion, and the amount of charge exchanged correlates directly to the surface density of separable surface groups \cite{mccarty_ionic_2007,mccarty_electrostatic_2008,lee_collisional_2018}. Alternative models exist for surfaces without separable surface molecules, which rely on aqueous ions in surface water films \cite{howell_dynamics_2001,schella_influence_2017} or in the atmosphere \cite{ducati_charge_2010}. 

Third, charging by transfer of material was observed for contacts involving polymers \cite{galembeck_friction_2014,burgo_triboelectricity:_2012,baytekin_mosaic_2011,baytekin_material_2012}. In such experiments, polymers were pressed into contact and material transfer can be verified in addition to charge transfer. Imaging the surfaces after separation with Kelvin Force Microscopy revealed a mosaic of positively and negatively charged microscopic spots \cite{baytekin_mosaic_2011}; the total net charge thus is the sum over many independent charge transfers.

Finally, the importance of polarization in generation of charge in granular media has been highlighted \cite{shinbrot_spontaneous_2008,yoshimatsu_self-charging_2017}. The charge separated in a contact depends on the field generated by all charges present in the surrounding, and minute initial charge on one of the surfaces may be amplified. 

The relevance of each of these four mechanisms, which may occur simultaneously in a single contact, and the extent of triboelectric charging during a contact will depend on the materials in contact and the aforementioned additionally relevant parameters and environmental conditions. The present knowledge of these mechanisms has been reviewed by several authors \cite{galembeck_friction_2014,williams_triboelectric_2012,lacks_contact_2011,matsusaka_triboelectric_2010,mccarty_ionic_2007}. 

Common to these four mechanisms is a {\it stochastic microscopic process.} The polarization mechanism may amplify some a priori unknown charge, but the electric field at the point of contact depends on a surrounding unknown charge landscape which justifies to assume a random electric field at the point of the contact zone. Also the microscopic processes at the contact of two bodies underlying the first three mechanisms can be modeled by two random surface distributions of donor and acceptor sites being pressed together \cite{apodaca_contact_2010}, where the donors and acceptors may represent trapped and valence band states, concentrations of separable surface groups, or concentrations of transferable polymer chains. Following this model, charge transfer is proportional to the overlap between acceptor and donor sites. The transferred net charge turns into a sum over random overlaps, and can be expected to be normally distributed in the central limit \cite{apodaca_contact_2010}.

A second group of studies has focused on the recombination of charges after the separation of the surfaces. The importance of the recombination and discharging of the surfaces to the full understanding of tribocharging has been discussed for long \cite{medley_frictional_1950,van_roggen_review_1967}. After all,
discharging in  spectacular sparks or lightnings is one of the most obvious manifestations of massive tribocharging.

Careful experiments have shown that even a single, nanoscale contact is followed by several discharging events \cite{horn_contact_1992,horn_contact_1993}. This can be understood by the fact that due to the limited surface conductivity a single discharging event cannot recombine the whole charge separated during the contact \cite{gibson_incendivity_1965,elsdon_contact_1976}. The superposition of the many discharging events then becomes similar to the discharging of a capacitor \cite{gibson_incendivity_1965,greason_triboelectrification_2013}. Moreover, experimentally observed decay times of triboelectric charges of tens of microseconds \cite{gibson_incendivity_1965} are comparable to estimated contact times for Hertzian collisions \cite{hertz:82}, suggesting that contact mechanics limit discharging times. 
 
Several mechanisms can be responsible for the individual discharge events, such as dark, glow and spark discharge \cite{gallo_corona-brief_1977}. Discharge by a spark discharge may cause the upper limit for the charge an insulator particle can carry after a collision \cite{matsuyama_charge_1995,matsusyama_impact_2006,mccarty_ionic_2007}. This threshold charge required to ignite a spark can be derived from Paschen's law \cite{paschen_ueber_1889}.

Which discharge mechanism will occur depends on parameters such as electric field strength, surface geometry, dielectric breakdown strength and separation velocity. However, all of these discharging processes through a gas have stochastic contributions like the probabilistic presence of ions formed by background radiation or illumination, stochastic collisional ionization of gas molecules and erratic path finding of streamers and sparks\cite{loeb_statistical_1948,davidson_statistics_1964}. In consequence, the realized conductance and the time the discharging persists will change stochastically.

If the contact involves granular particles, additional parameters such as surface roughness, particle shape variations or rotary particle motion will result in strong fluctuations of factors such as surface geometry and separation velocity. The net attenuation given by conductance and contact mechanics thus can be expected to be a random variable.

We conclude that both the mechanisms associated with triboelectric charging and subsequent discharging are of stochastic nature. The statistics of both charging and discharging and the combined effect have rarely been discussed. Fluctuations of the net charge of individual particles have been observed in previous studies \cite{mccarty_ionic_2007,waitukaitis_size-dependent_2014,yoshida_estimation_2003,mazumder_measurement_1991,matsuyama_impact_2003}. In one of these studies a non-normal distribution of the generated charge is reported \cite{mccarty_ionic_2007}. This nontrivial property motivates a closer investigation of the charge distribution and the relation to the two underlying mechanisms.


\section{\label{sec:exp} Experimental methods and results}

Stochastic triboelectric charging implies fluctuations of the charge on granular particles generated in identical configurations. We quantify this by repeated measurements of the charge generated in a single contact or during multiple collisions of a granular particle. A key aspect of our experiment is the minimization of charges generated during handling of the particles prior to the single contact measurement. Handling is inevitably connected to forming and breaking of contacts and as such creates charges on the particles which will be superimposed to the charge generated in the contact to be tested. Another elegant approach to minimize charging during handling is presented in a recent study, where the particles are levitated in an acoustic levitator prior to contact charging \cite{lee_collisional_2018}. 

\subsection{\label{sec:expI} Single contacts}

\begin{figure}
\includegraphics[width=0.95\linewidth]{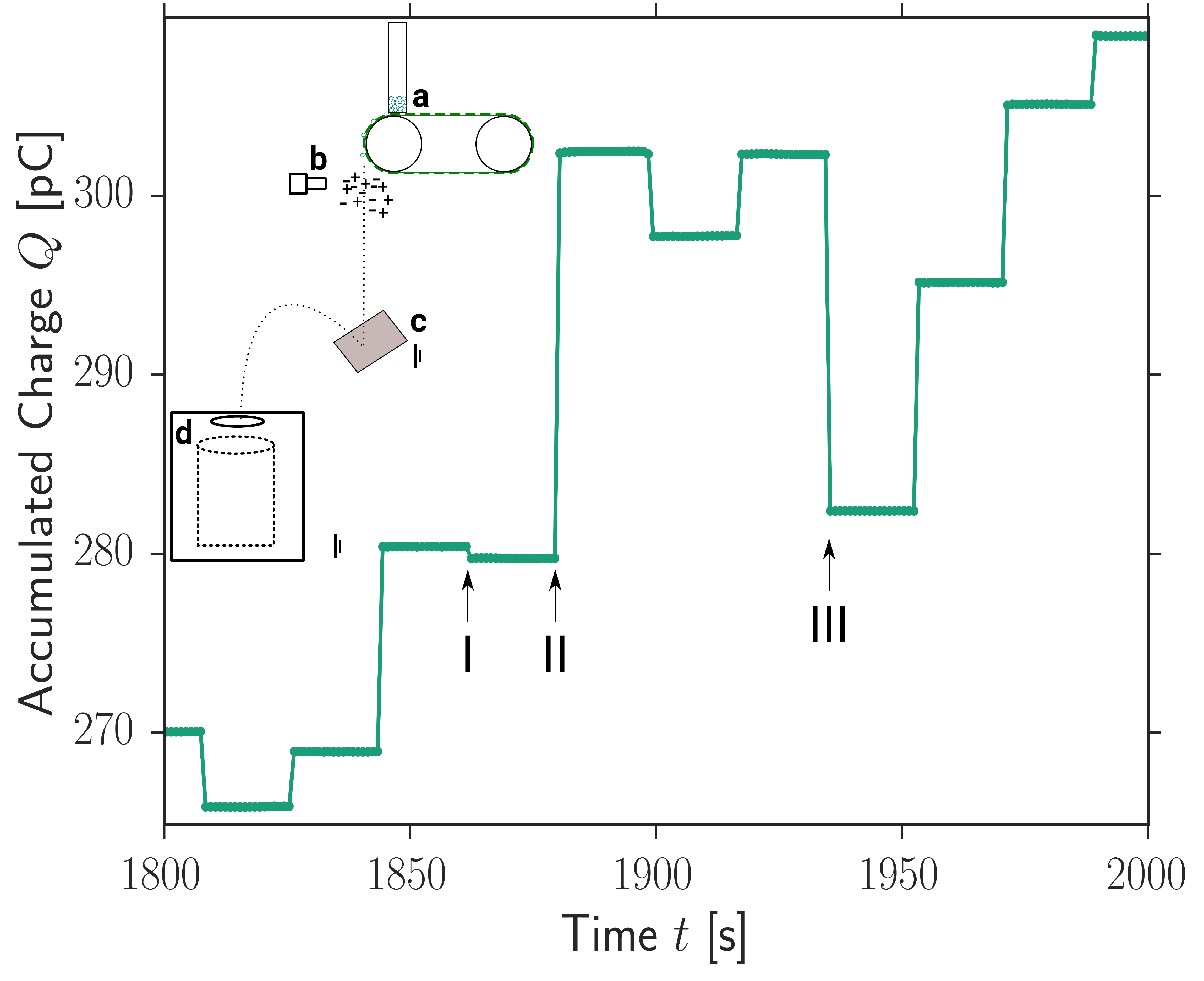}
\caption{Detail of the charge measurements, here glass beads bouncing off a PTFE slab. Each particle dropping into the Faraday cup causes a change in charge accumulated in the Faraday cup, resulting in a jump of the curve. Steps I, II, and III exemplify the large variability in magnitude and sign of the charge on the particles. The inset shows the measurement setup, with the particle dispenser (a) dropping individual particles through a charge-neutralizing cloud of ionized air created by an ionization needle (b) onto a collision target (c). The particles bounce after a single collision into a Faraday cup inside of grounded conductive housing (d).}
\label{fig:charge}
\end{figure}

A schematic of the measurement setup is shown in the inset of Figure~\ref{fig:charge}. Spherical particles (soda-lime glass beads, 4\,mm diameter) are released from the reservoir one at a time through use of a particle dispenser (a). The dispenser picks particles by rotating a wheel with dimples below the reservoir. The particles are released from the dimples with rotation of the dispenser wheel and fall an identical distance of 300\,mm. The particles are discharged while passing through ionized air with positive and negative ions created by an ionization needle (Haug OPI) (b). While the kinetic energy of the particles is given by their falling height, the release from the dispenser wheel imparts them  with an additional unknown rotational component. The particles hit a collision target at an angle of 60$^{\circ}$ (grounded copper slab or polytetrafluoroethylene (PTFE) slab,(c)) and fall into the Faraday cup (d) where their charge is measured using a Keithley 6514 electrometer.

The electrometer measures continuously  the charge accumulated by the Faraday cup. The particles falling into the cup led to equidistant changes in the charge (see exemplary charge curve in Fig.~\ref{fig:charge}). As can be seen from the highlighted examples (I, II, III), the charges accumulated during the collision by the glass beads can vary orders of magnitude and also in their sign.
The charge distribution $P(Q_n)$ of the net particle charges $Q_n$ is determined by counting each change of charge above a threshold of 0.1\,pC. This threshold  is necessary to take the drift of the electrometer into account. 

All particles used in our experiments are first rinsed with water and ethanol and then cleaned in an Argon plasma for 10 min (Diener electronic Femto plasma cleaner). Between 500 to 1500 particles were dropped for each measurement, the exact numbers can be found in 
table \ref{tab:moments}.

\begin{figure}
\includegraphics[width=0.95\linewidth]{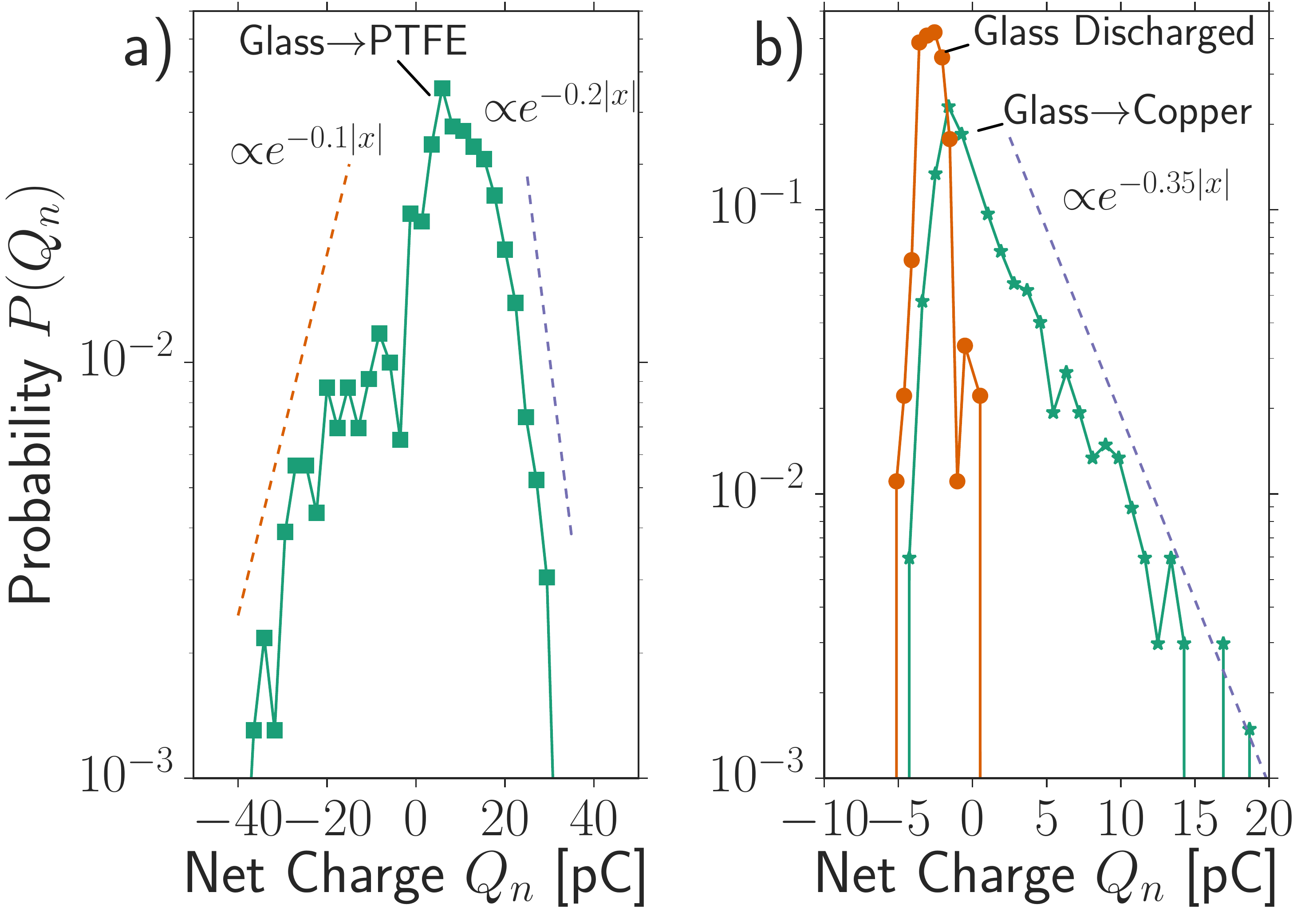}
\caption{Probability distribution for charges accumulated in single collisions of Glass beads with a PTFE slab (a) and of Glass beads with a Copper slab (b). The dashed lines act as a guide to the eye and show the slope of the exponential decay of the tails. Also shown in (b) is the distribution of charges on the particles after passing the ionization needle without colliding with a target.}
\label{fig:sc_dist}
\end{figure}

In order to identify our background, we first measure the charge distribution of spheres which have not collided with a target, i.e.~dropped directly into the Faraday cup. Without the ionization needle, these particles accumulate between 100\,pc and 1\,nC on their surface. After including the ionization needle in the setup, the residual charge is reduced to a narrow distribution between 1\,pC and -5\,pC with a mean of -2.65\,pC; this distribution is shown in figure \ref{fig:sc_dist}, b. 

Introducing now a collision target in the path of the particles changes the charge measured on the particles (see Fig.~\ref{fig:sc_dist}). Instead of gathering a fixed amount of charge during the collision, the particle charge becomes wider distributed, with a strong dependency on the collision target material. The particles accumulate a mean charge of 4.12\,pC in the case of of glass beads bouncing off a PTFE slab (Fig.~\ref{fig:sc_dist}, a), while after collision with copper a mean charge of 0.90\,pC is accumulated (Fig.~\ref{fig:sc_dist}, b). The distributions are asymmetric, with a skewness of -1.16 (PTFE) /  1.89 (Copper), and are fat-tailed with an excess kurtosis of 1.63 (PTFE) / 5.20 (Copper). The tails of the distributions decay approximately exponentially. Noteworthy is the pronounced presence of charges of both signs. 

An average charge densities of roughly 100$e^-/\mu m^2$ for both cases can be estimated from the mean net charge on the particles of 0.90\,pC (Copper) / 4.12\,pC (PTFE), assuming a Hertzian contact among a sphere and a flat surface \cite{hertz:82} and taking into account a falling height of 30\,cm. This is comparable to previous studies on triboelectricity, where numbers of 300$e^-/\mu m^2$ \cite{harper_electrification_1961} or 500$e^-/\mu m^2$ \cite{mccarty_electrostatic_2008} are reported. However much higher charge densities above 1200$e^-/\mu m^2$, more than ten times the mean, occurred on 3.5\% of the spheres, exemplifying the effect of skewness and fat-tailedness of the measured distributions.

\subsection{\label{sec:expII} Multiple contacts}

\begin{figure}
\includegraphics[width=0.95\linewidth]{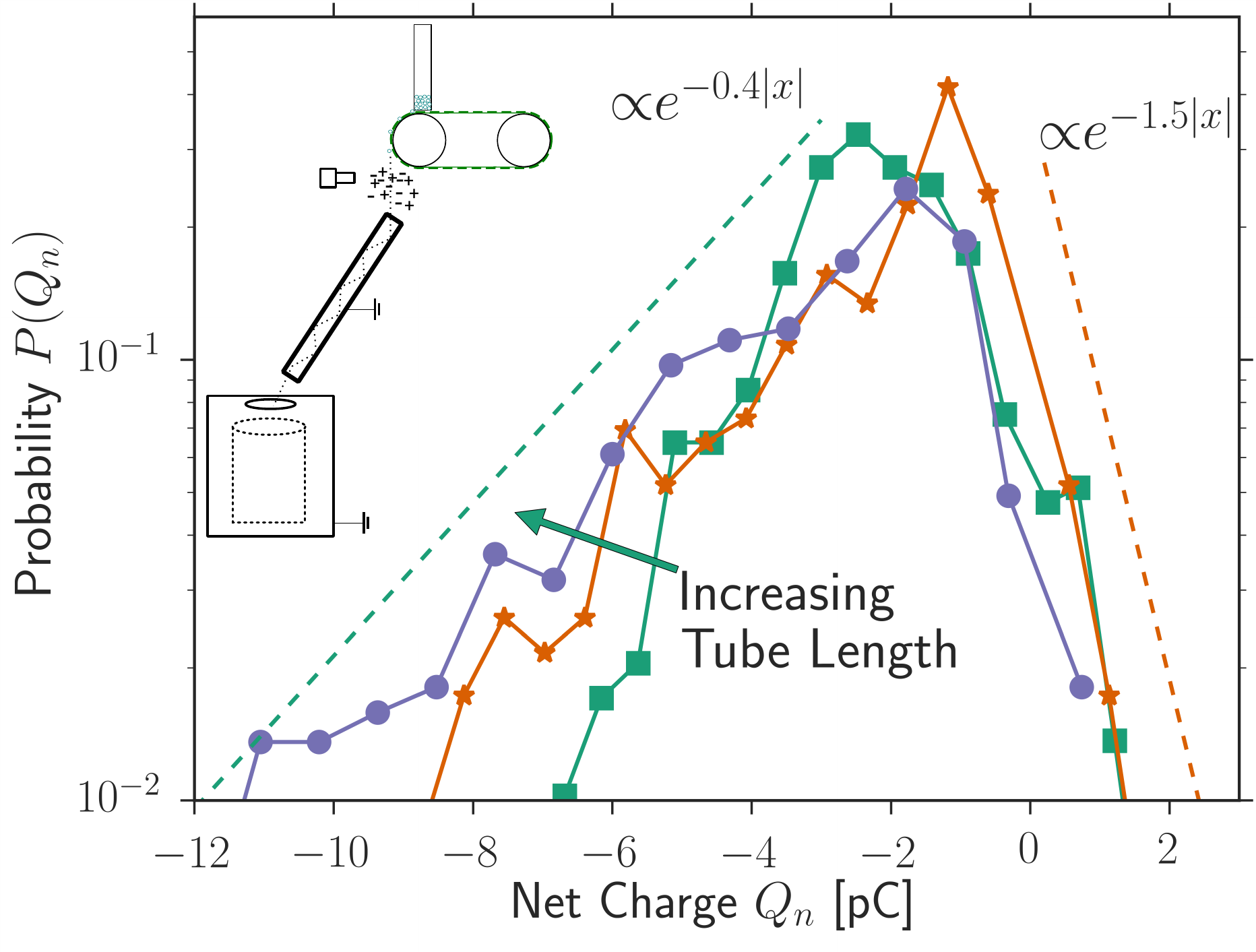}
\caption{Probability distributions for charges accumulated by individual glass beads bouncing down inclined PMMA tubes with lengths of 20\,cm, 80\,cm, and 100\,cm (as indicated by the arrow). The dashed lines are guides to the eye and highlight the approximately exponential tails of the distributions. Several key features like the asymmetry and the approximately exponential decay of the probability distributions do not change with number of collisions. The inset shows the measuring setup for multiple collision measurements.}
\label{fig:tube}
\end{figure}

In a second series of experiments we replace the large glass spheres by smaller spheres with 500-560\,$\mu$m diameter; these particles are made of either soda lime glass (Worf Glaskugeln GmbH) or polystyrene (Spheromers CS 500, Microbeads) and were cleaned as before. With these small spheres a charge neutralization down to a residual background charge between -0.1\,pC and +0.1\,pC is achieved, a much better value than for the larger spheres. The collision targets are replaced by tubes oriented at a 45$^{\circ}$ angle, such that the particles perform many contacts inside the tube before falling into the Faraday cup (see inset Fig.~\ref{fig:tube}). Some experiments are performed inside of a climate chamber to test the influence of ambient conditions such as the relative air humidity (RH). 

This setup  allows us to test how the charge distribution changes as a function of the average number of contacts a particle experiences. For this purpose we vary the length of the polymethyl methacrylate (PMMA). We note that the contact mode, the ratio of normal and tangential component in a contact, may change along with the number of contacts. The resulting charge distributions are shown in Fig.~\ref{fig:tube}. The number of contacts will grow systematically with tube length, but also the ratio of tangential to normal force component in each contact will change along the tube. The measured average charge grows with tube length (from -2.37~pC to -3.55~pC), accompanied by an increasing negative skew (from -0.34 to -1.38). The features mentioned for the charge distributions measured after a single contact also hold for multiple contacts. All distributions possess approximately exponential tails and a strong asymmetry and have a positive excess kurtosis (from 0.72 to 1.96).

\begin{figure}
\includegraphics[width=0.95\linewidth]{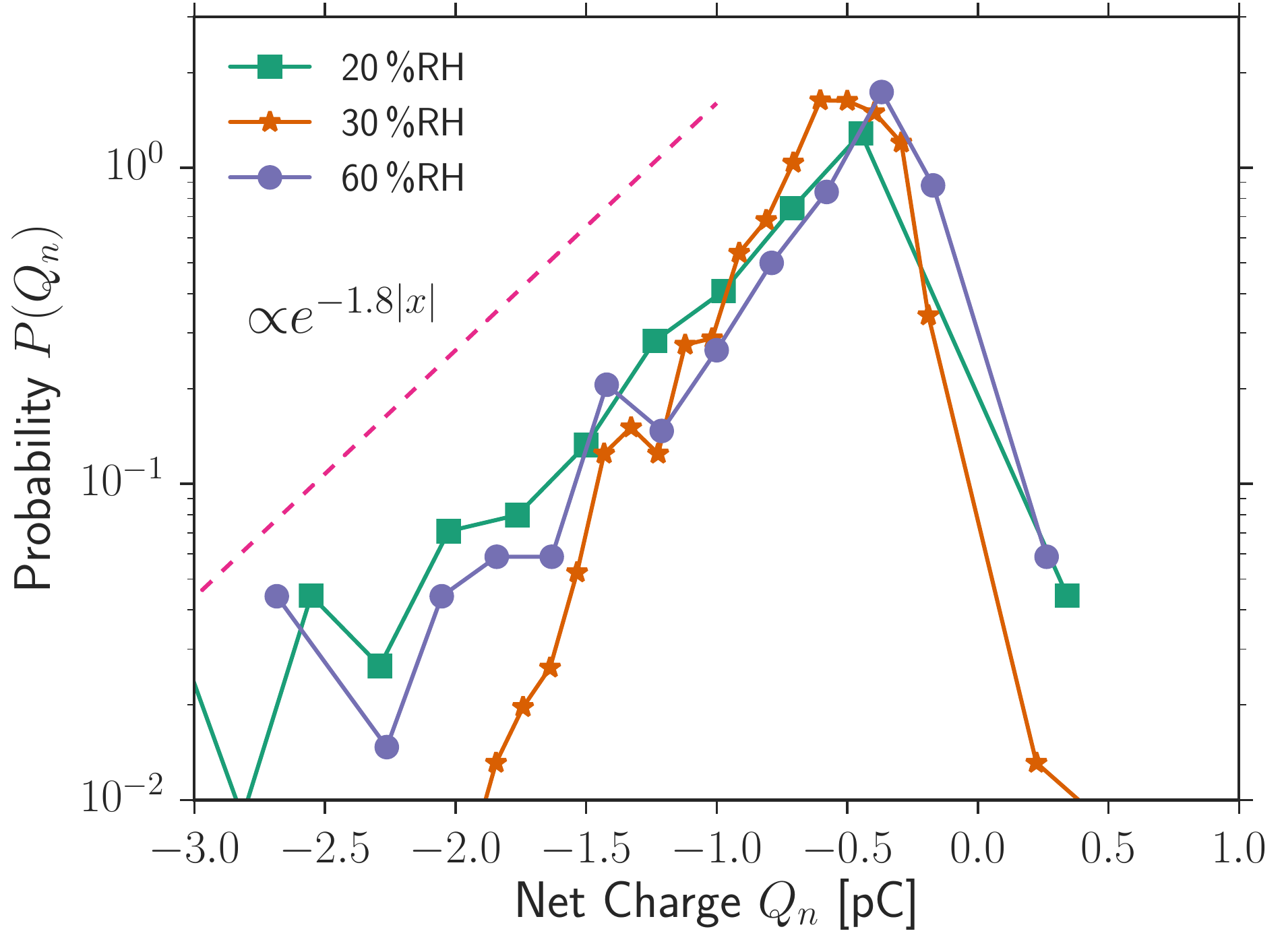}
\caption{The charge distributions of glass beads bouncing down a stainless steel tube of 40 cm length, measured at different relative air humidities (RH). The distributions show no systematic trend with the relative humidity. Again we observe wide, asymmetric distributions.}
\label{fig:RH}
\end{figure}

Because  environmental conditions influence triboelectricity, we perform additional experiments where we vary the relative air humidity using the climate chamber described in \cite{schella_influence_2017}. The samples and the setup are kept at constant conditions for at least half an hour before the start of the measurements. The variation in relative humidity during the measurement time is smaller than $\pm$ 8\%RH. The charge distributions measured at 20\,\%, 30\,\%, and 60\,\%RH are displayed in Fig.~\ref{fig:RH}. Again we observe wide, asymmetric distributions irrespective of ambient humidity; the mean values, skew, and kurtosis are listed in table \ref{tab:moments}. A decrease of the mean charge with relative humidity can be observed. We also observe an enhanced drift of the charge measurements at higher humidities, similar to previous studies \cite{ducati_charge_2010}. This effect prevents charge measurements above 60\,\%RH.

\begin{figure}
\includegraphics[width=0.95\linewidth]{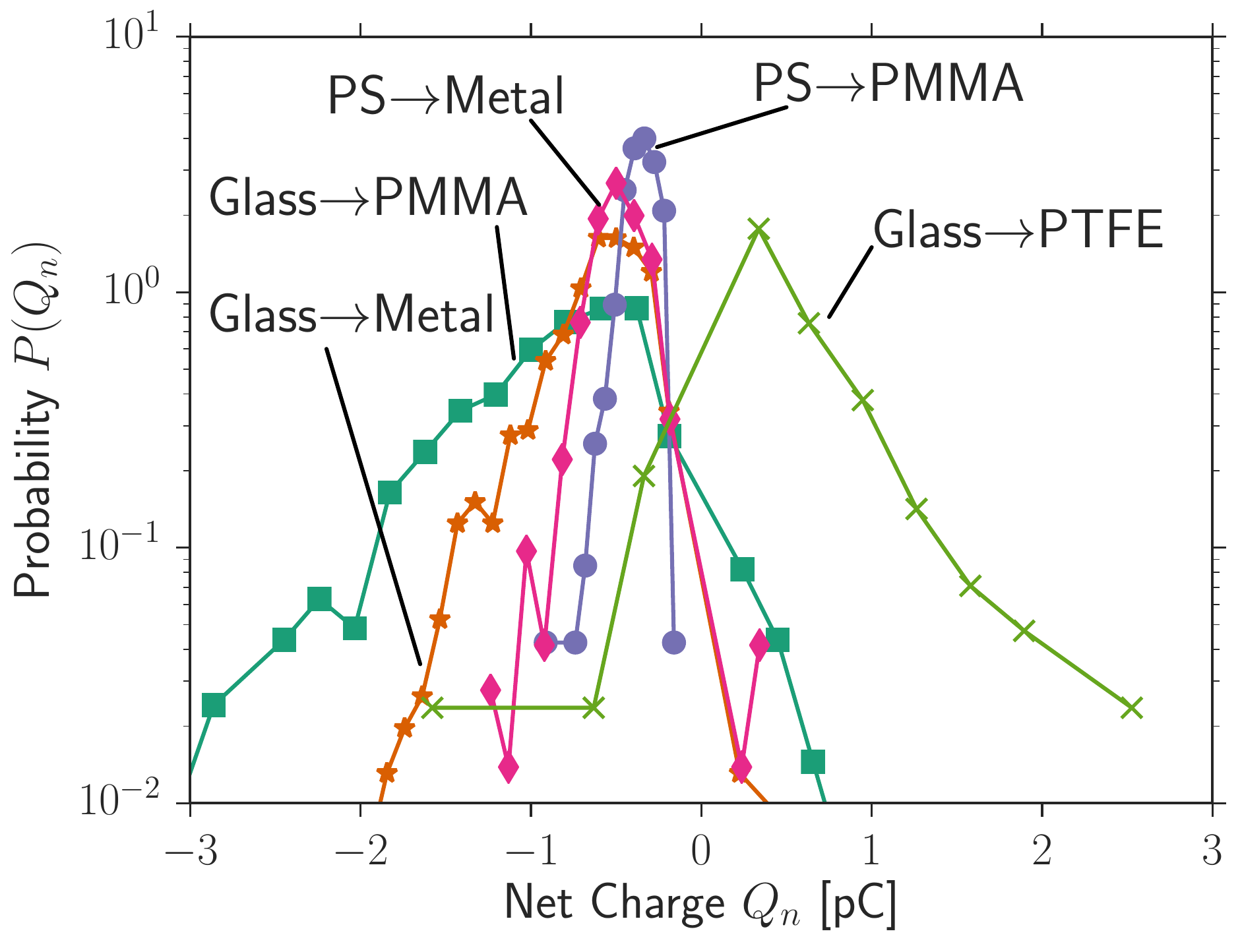}
\caption{Probability distributions for the charges accumulated by individual particles bouncing down tubes made of different material. The particle materials glass and PS were combined with tube materials of PMMA, grounded stainless steel and PTFE. The general appearance of the distributions does not depend on the specific material combination, all combinations lead to sharp, fat-tailed distributions with an excess kurtosis ranging from 2.03 to 4.69. }
\label{fig:matcomb}
\end{figure}

In order to test how generic the above described features of the charge distributions are, we tested a number of additional material combinations using both glass and polystyrene (PS) spheres, and tubes made either from PMMA, grounded stainless steel, or PTFE. The resulting charge distributions are displayed in Fig.~\ref{fig:matcomb}. All distributions are asymmetric with skew values with the same sign as the mean. And they are fat-tailed with an excess kurtosis ranging from 2.03 to 4.69. Thus the characteristic shape of the distributions obtained in the previous measurements is preserved, although the average charge varies from  -0.87\,pC (glass against PMMA) to 0.52\,pC (glass against PTFE). We also considered studying same-material tribocharging using 0.5\,mm PMMA spheres in the PMMA tube. The achieved charges were typically less than 0.1\,pC, not enough to be reliably measured.

\input{table.tex}

\begin{figure}
\includegraphics[width=0.95\linewidth]{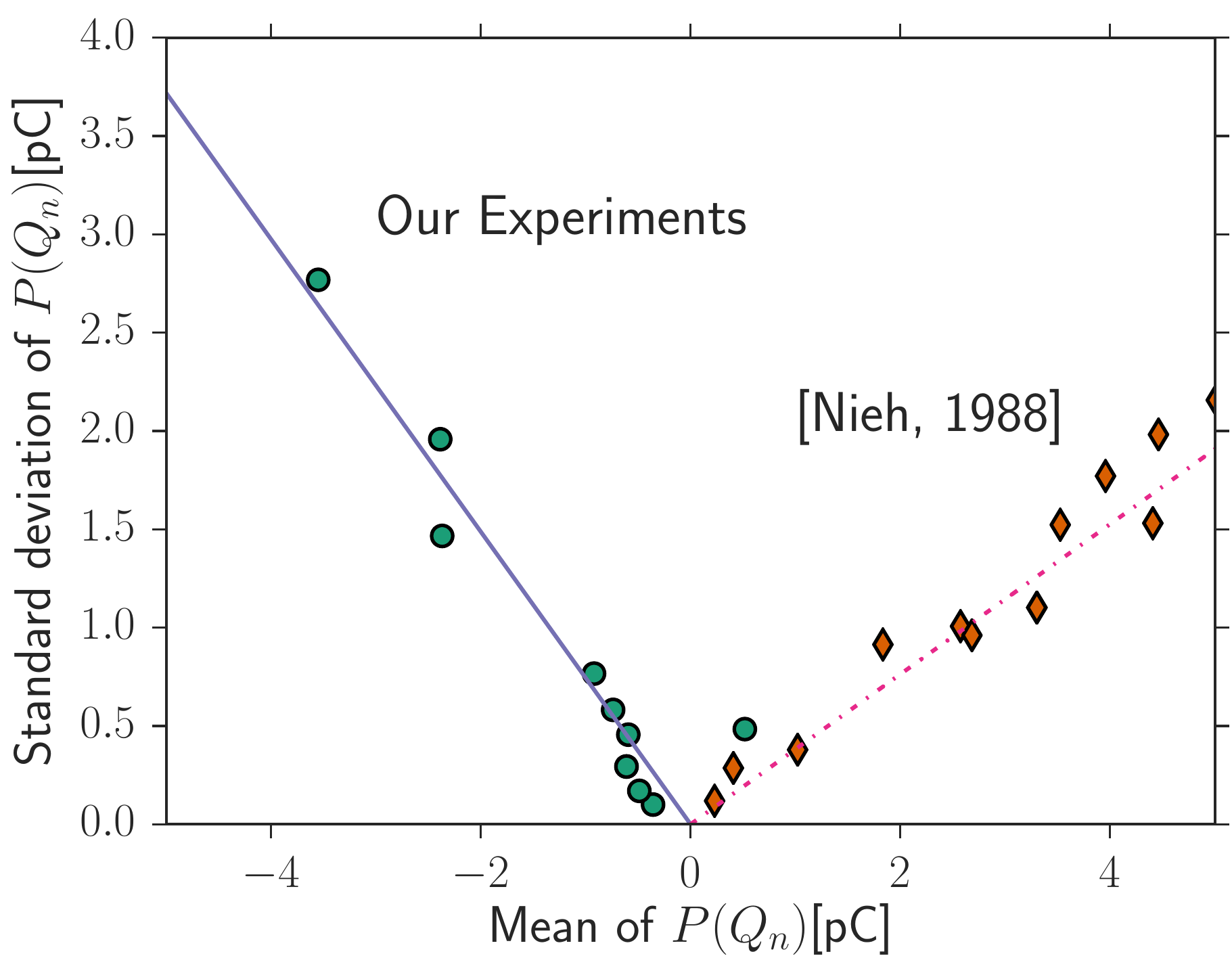}
\caption{Standard deviations $\sigma_{n}$ of the charge distributions $P(Q_n)$ plotted as a function of the respective means $\mu_{n}$. The standard deviation grows linearly with the mean charge. Circles represent data from the measurements presented here, diamonds are values taken from a previous study \cite{nieh_effects_1988}. The linear approximations describe  measurements with a wide range of parameters, i.e.~different materials, different humidities, different number of contacts, and different setups.}
\label{fig:fits}
\end{figure}

The moments of all charge distributions reported here can be found in table~\ref{tab:moments}. None of these distributions is close to a normal distribution. Beyond the peculiar common shape of the charge distributions, they share another feature: the standard deviation $\sigma_{n}$ and the mean $\mu_{n}$ of the charge distribution seem to be correlated. Figure \ref{fig:fits} displays $\sigma_{n}$ as a function of $\mu_{n}$. The data points are taken from both our measurements with the 500\,$\mu$m particles and  from a previous study, where the standard deviation and the mean were reported \cite{nieh_effects_1988}. A linear correlation of the width to the average charge can be observed in both cases, only the slopes differ for the two data sets. A correlation thus is present irrespective of material combination, average number of collisions, ambient conditions, setup and other experimental conditions.

\section{\label{sec:theoryII} Double stochastic model} 
Our measurements confirm that the charge generated after breaking a contact between a particle and a solid is not a material constant, but a random variable. Particles colliding with the same material may gather positive as well as negative charges. The measured distributions of the triboelectric charges are asymmetric, i.e.~mean and median differ, and they also exhibit long, close to exponentially decaying tails. These qualitative properties of the charge distributions do not change with relative humidity, type of material, or the average number or mode of the collisions. Moreover, the width of the distributions increases with the mean charge. This means that increasing the charge on a particle by more contacts or specific material combinations will also increase the fluctuations of that charge. 

The observed charge distributions $P(Q_{n})$ thus considerably differ from normal distributions. Availability of an improved functional form based on the observations would allow using realistic charge distributions when modeling granular media, and may guide development of extended microscopic theories. We rationalize a model based on the concepts summarized in Sec.~\ref{sec:theoryI}. It can be seen that triboelectric charging is dominated by two processes, charge separation and charge recombination.

It is presently difficult to derive a functional form for the charge distribution predicted by the charge separation caused polarization and induced charging. However, the three other discussed fundamental mechanisms strongly suggest a normal distribution of initially separated charges $Q_{c}$ \cite{apodaca_contact_2010}. As discussed in Sec.~\ref{sec:theoryI}, the net charge on a particle is the sum over many donor and acceptor sites, which may represent densities of trapped and valence band states, concentrations of separable surface groups, or concentrations of transferable polymer chains. A quantitative estimation can be derived from observed sizes of such sites \cite{baytekin_mosaic_2011}. Sites are observed on length scales of 4.5$\mu m$ and 0.44$\mu m$. In our experiment we can estimate a Hertzian contact radius of 127$\mu m$ for the 4~mm glass spheres contacting the copper plate in the single contact experiments. From this one can estimate a lower limit for the number of involved sites of about 2500, justifying the use of the central limit theorem and the approximation of a normal distribution for the net separated charge $Q_{c}$.

The separated charges tend to equilibrate. This equilibration may happen by visible spark discharging, once the Paschen limit for initiating a breakdown is overcome \cite{matsuyama_charge_1995}. Below the Paschen limit, discharging occurs by other gas discharging mechanism like dark and glow discharge \cite{gallo_corona-brief_1977}. The kinetics of the discharging of insulator surfaces in the presence of various discharging mechanisms and the distribution of the exchanged charges are hard to be derived. Experimental observations have yet shown, that many discharging events superpose to the exponential discharging kinetics of a capacitor \cite{gibson_incendivity_1965}. The attenuation $\alpha_{d}$ realized by discharging after a contact depends on the conductivity realized by the particular discharging mechanism and the time the two involved surfaces stay in proximity where discharging is efficiently possible. The total discharging of two surfaces intrinsically is the sum over several discharging events \cite{gibson_incendivity_1965,horn_contact_1992,horn_contact_1993}, and a normal distribution of the attenuation $\alpha_{d}$ is suggested as an approximation.

The net charge $Q_{n}$ remaining after a particular contact thus can be assumed to emerge from a normally distributed initially separated charge $Q_{c}$, which has decayed exponentially with a normally distributed attenuation $\alpha_{d}$: 
\begin{equation}
	Q_n = (\left. Q_{c}\right|\mu_{d}, \sigma_{d}) \cdot\exp{(-(\left.\alpha_{d}\right|\mu_{d}, \sigma_{d}))},
\end{equation}
with the means and standard deviations $\mu_{c}$, $\sigma_{c}$, $\mu_{d}$ and $\sigma_{d}$. The exponential of a normally distributed variable $\alpha_{d}$ itself represents a lognormally distributed variable. The net charge $Q_{n}$ consequently is the product of a normally and a lognormally distributed variable, i.e. has a normal-lognormal distribution $P(Q_{n})$~\cite{yang_normal_2008}. By writing 
\begin{equation}
	\exp{(-(\left. \alpha_{d}\right|\mu_{d}, \sigma_{d}))} = \exp{(-(\left. \alpha_{d}\right|0, \sigma_{d}))} \cdot \exp{(\mu_{d})}.
\end{equation}
and multiplying the new exponential term to the normally distributed variable $Q_{c}$ one obtains
\begin{equation}
	P(Q_{n}) = P((\left. Q_{c}\right|\tilde{\mu}_{c}, \tilde{\sigma}_{c}) \cdot  \exp{(-(\left. \alpha\right|0, \sigma_{d}))}),
\end{equation}
a function of only three parameters $\tilde{\mu}_{c}$, $\tilde{\sigma}_{c}$ and $\sigma_{d}$. These parameters incorporate all material and ambient parameters of the configuration relevant to the charge separation and the recombination processes. Presently, in the absence of a microscopic theory, these are phenomenological parameters.

\begin{figure}
\includegraphics[width=0.95\linewidth]{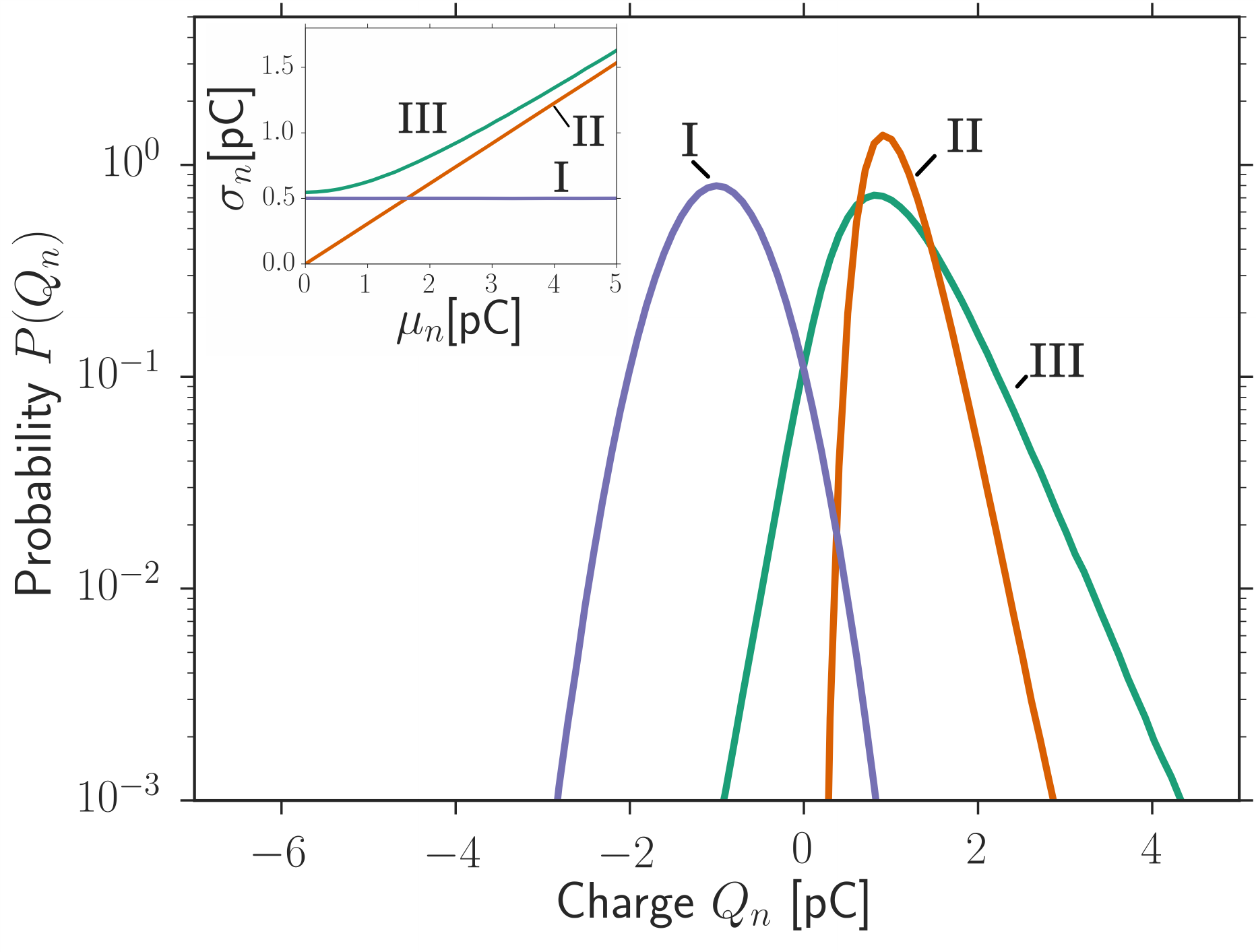}
\caption{Expected shapes of the probability distribution $P(Q_n)$ of the net charges $Q_n$. I: For $\sigma_{d}\rightarrow 0$ no discharging occurs. The distribution is the normal distribution of the initially separated charges (here: $\tilde{\mu}_{c}=-1,\tilde{\sigma}_{c}=0.5,\sigma_{d}=10^{-6}$). II: The distribution turns into a lognormal distribution if the charge separation is deterministic and $\tilde{\sigma}_{c}\rightarrow 0$ (here: $\tilde{\mu}_{c}=1,\tilde{\sigma}_{c}=10^{-6},\sigma_{d}=0.3$). III:  The intermediate case shows exponential tails, asymmetry about the mean and both positive and negative values can be reached (here: $\tilde{\mu}_{c}=1,\tilde{\sigma}_{c}=0.5,\sigma_{d}=0.3$).
The inset shows how the the standard deviation $\sigma_{n}$ depends on mean $\mu_{n}$ of the net charge $Q_n$, starting from each of the three cases above and varying the mean exchanged charge $\tilde{\mu}_{c}$. In the case of a lognormal and the normal-lognormal distribution a linear relation emerges.}
\label{fig:model}
\end{figure}

The analytical handling of this normal-lognormal distribution $P(Q_{n})$ is difficult \cite{yang_normal_2008}. We therefore model $P(Q_{n})$  numerically by drawing normally distributed random variables for the $Q_{c}$ and $\alpha_{d}$ and calculating the histogram for the final charge $Q_{n}$. It is instructive to consider the limiting cases of the predicted normal-lognormal distribution for $Q_{n}$ displayed in Fig.~\ref{fig:model}. 

$P(Q_{n})$ approaches a normal distribution and $Q_n=Q_c$  with $\sigma_{d}\rightarrow 0$, (Fig.~\ref{fig:model}, I). For a normal distribution of $Q_n$ charges of both signs are possible, the skewness and the kurtosis vanish, and the tails of the distribution decay faster than exponential, as can be seen from the semilogarithmic plot.

A situation with $\tilde{\sigma}_{c} \rightarrow 0$, i.e.~a deterministic $Q_{c}=\mu_{c}$, results in lognormally distributed $Q_n$ (Fig.~\ref{fig:model}, II). In this case the skewness and the excess kurtosis do not vanish, but the $Q_n$ all have the same sign.

The intermediate normal-lognormal cases combining non-vanishing $\sigma_{d}$ and $\tilde{\sigma}_{c}$ result in distributions for $Q_n$ that are not symmetric about the mean, with finite skewness and excess kurtosis, with net charges of both signs possible and with approximately exponentially decaying probabilities (Fig.~\ref{fig:model}, III). The probability distribution of the net charge in this normal-lognormal case thus deviates qualitatively from both the normal distribution and the lognormal distribution.

The inset of Fig.~\ref{fig:model} shows the relations between the  mean $\mu_{n}$ and the standard deviation $\sigma_{n}$ for all three cases. The intermediate case with a normal-lognormal distributed net charge $Q_{n}$ results in a linear relationship, at least for larger values of  $\mu_{n}$. This linear dependence exists also in the case of a lognormal distribution \cite{limpert_log-normal_2001}. The slopes of the linear regimes of the lognormal and the normal-lognormal distributions depend on $\sigma_{d}$, thus are characteristic for the discharging mechanism in the respective situation.

The normal-lognormal distribution thus exhibits the characteristics of the charge distributions observed in the experiments. Additionally it can be motivated from observations on the individual steps of charge separation and recombination presented in the literature. We consequently propose this distribution with three parameters as a minimum model to describe the shape of charge distributions generated by triboelectric charging.


\section{\label{sec:disc} Discussion}

The measurements presented in sec.~\ref{sec:exp} demonstrate that the distribution of charges generated during the collisions of insulating granular particles posses characteristic features which are independent of the specifics of the experiment. These features include the possibility of charges of both signs, asymmetry, close to exponentially decaying tails and a linear correlation among standard deviation and mean. 

The particular shape of the charge distribution proves that the mean particle charge is not sufficient to correctly model particle charging. Particles with charges of both signs are possible for the same material combination, and due to the approximately exponential decay of the distribution particles with extreme charges are more likely than expected from e.g. a normal distribution. The mean charge does not even describe the most likely charge that a particle has due to the strong asymmetry of the charge distribution.

The correlation among width and mean also implies, that parameters like the material combination or the number of collisions simultaneously determine how much charge is separated on average and how wide the distribution of separated charges is. Situations with large net charges will also have the greatest variation.

It is possible to systematically vary the mean net charge by using different material combinations or increasing the number of contacts. However, the charge distributions keep their characteristic general shape.  Especially, they do not converge towards a normal distribution when the number of contact events is increased (controlled by the tube length). This implies that subsequent contacts of the particles do not fulfill the main requirements of the central limit theorem: statistical independence. In practical terms, this statistical dependence of subsequent triboelectric charging events demands a high efficiency of the initial neutralization of the particles used in experiments. Otherwise any residual charges from handling the particles will bias the results.

It is unclear whether this statistical dependence is due to the charging or the discharging process. Charges present on the particles may affect the uptake of new charges \cite{shinbrot_spontaneous_2008} as well as the discharging process \cite{stollenwerk_spatially_2007}. In a recent study on the collisional triboelectric charging of a single sphere a linear increase of the charge on the sphere with the number of collisions was observed \cite{lee_collisional_2018}, suggesting a statistical independence of the charges generated during individual collisions.  However, the average charge generated during collisions was orders of magnitude smaller than in the experiments presented here. In the same study a strong dependence of the transfered charge on electrical fields is reported. In another recent study on polymer particles saturation of charge is apparent after tens of collisions \cite{chowdhury_charge_2018}. Both these results suggest that a certain level of charge on the particles is required to obtain statistical dependence between charges transfered in subsequent collisions. 

An alternative hypothesis for the statistical dependence observed in our experiments might be the small surface area of the particles. It cannot be excluded that during motion in the tube the particles contact the tube with the same spot several times or even switch to sliding motion. Depletion of charge carriers in this spot could cause statistical dependence. 

We suggest a three-parameter normal-lognormal distribution for the net charge in sec.~\ref{sec:theoryII}. This distribution reproduces the characteristic features of the experimentally measured charge distributions. We rationalized this shape of the distributions by combining two stochastic processes, charge separation during contact and subsequent charge recombination. This suggests that a complete description of triboelectricity requires simultaneous understanding of both the mechanisms relevant for charge separation and the mechanisms relevant for charge recombination. 

The complexities arising from the combination of both the mechanisms can be illustrated by trying to connect the three parameters $\tilde{\mu}_{c}$, $\tilde{\sigma}_{c}$ and $\sigma_{d}$ (or $\mu_{c}$, $\sigma_{c}$, $\mu_{d}$ and $\sigma_{d}$) with the microphysical processes and the relevant parameters discussed in Sec.~\ref{sec:theoryI}. Humidity for example is expected to increase mobility of charge carriers, in particular enhances presence and mobility of ions on solid surfaces. Humidity consequently could be expected to increase the mean exchanged charge $\mu_{c}$ and its variation \cite{pence_effect_1994}. On the other hand, humidity increases the mobility of charge carriers and the conductivity of air, thus can be expected to enlarge $\mu_{d}$ and to minimize the residual net charge \cite{biegaj_surface_2017,schella_influence_2017}. The factors which determine the dominating influence may be very subtle.

In future studies either charging or discharging shall be addressed separately or time-resolved measurements should be made. Humidity can be expected to affect both charging and discharging, but other parameter might be identifiable, which affect selectively the discharging. Gas atmospheres with varying conductivity or breakdown threshold with constant humidity, like SF$_6$-containing gas, may make isolating charging statistics possible \cite{matsuyama_charge_1995,mccarty_ionic_2007}. A setup to achieve situations with suppressed discharging using a bias voltage has also been proposed \cite{Ireland_contact_2009}. A situation where discharging is increasingly suppressed should converge to normally distributed charges $Q_n=Q_c$ with net charges of both signs possible (Fig.~\ref{fig:model}, I). In such a situation $\mu_{c}$ and $\sigma_{c}$ of the charging step may be studied. Time resolved measurements also may allow to distinguish charging and subsequent discharging and to follow the discharging kinetics~\cite{gibson_incendivity_1965,horn_contact_1992,horn_contact_1993}. 

The intermediate cases, where stochastic charging and stochastic discharging are combined, result in normal-lognormal distributions. The parameters $\mu_{c}$, $\sigma_{c}$, and $\mu_{d}$ turn into only two independent parameters $\tilde{\mu}_{c}$ and $\tilde{\sigma}_{c}$ (see derivation in Sec.~\ref{sec:theoryII}). In this case the linear correlation among the mean $\mu_{n}$ and the standard deviation $\sigma_{n}$ of the net charge that we found in our experiments and also in a previous study reporting these parameters is particularly interesting. The model predicts that the slopes of the relation among $\mu_{n}$ and $\sigma_{n}$ are characteristic of $\sigma_{d}$, the standard deviation of the distributions of the attenuation coefficients $\alpha_d$. Such behavior may enable disentangling the contributions from the charging and the discharging to the final net charge in further studies. 

The reasoning of the normal-lognormal distribution by combining a normal distribution for the charging step and a normal distribution for the attenuation coefficient may, of course, be a first order approximation. It relies on the experimental observations of a surface mosaic of separated charges, and an exponential decay of the charge. While the normal distribution of the separated charges is well justified by the central limit, the distribution of the attenuation coefficient is not yet clarified. Situations may arise, where several charge separation mechanisms or different charge dissipation mechanisms act simultaneously and may alter the distributions. Still, all tested situations in this work, including various materials, contact number and ambient conditions, match the predictions by a normal-lognormal distribution. This suggests that the proposed first order approximation covers already many aspects of triboelectric charging, and justifies a view of triboelectric charging as a combination of two coupled stochastic processes.

\section{\label{sec:conc}Conclusion}
Triboelectric charging of granular particles is a stochastic process. The statistics of the generated net charges on the particles reveal several distinctive features:  the distributions are more heavy-tailed than normal distributions with an exponential decay of the probability, are asymmetric, exhibit charges of both signs and exhibit a linear correlation among width and mean. We show that a normal-lognormal distribution is compatible with the observations. We rationalize the normal-lognormal distribution by describing triboelectric charging as a two-step random process. In the first step the charges are separated during contact, while in the second step charge recombination occurs after separation of contact. Moreover we find that subsequent charging events are not statistically independent as the distribution of multiple triboelectric charging events does not converge to a normal distribution.

\section*{Acknowledgments}
The authors would like to thank Masato Adachi for proofreading the manuscript. Funding provided by the German Science Foundation under DFG BO 4174/2-1, SP 714/12-1, and Cluster of Excellence Engineering of Advanced Materials is gratefully acknowledged.



\bibliography{main} 

\end{document}

%% file: table.tex
\begingroup
\begin{table*}[htbp]
\caption{Measured charge distributions and their moments.}
\label{tab:moments}
\begin{tabularx}{\textwidth}{XX}

\begin{tabular}{p{.8cm} p{1.1cm} p{1.2cm} p{1.2cm} p{1cm} p{1.3cm} | p{1cm} p{1.2cm} p{1.2cm} p{1.1cm} p{1.2cm} p{1cm}}
\hline\hline
Fig.	&	Sphere\newline Material	&	Sphere\newline Diameter	&	Target Material	&	Length	&	Relative\newline Humidity	&	Mean\newline Charge	&	Median Charge	&	Standard\newline Deviation	&	Skew	&	Excess\newline Kurtosis	&	Sample Size	\\
	&		&	 [mm]\vspace{.05cm}	&		&	[cm]	&	[\%]	&	[pC]	&	[pC]	&	[pC]	&		&		&		\\\hline
\ref{fig:sc_dist} a)	&	Glass	&	4	&	PTFE	&	Plate	&	50	&	4.12	&	6.83	&	14.54	&	-1.16	&	1.63	&	991	\\
\ref{fig:sc_dist} b)	&	Glass	&	4	&	Copper	&	Plate	&	50	&	0.90	&	-0.49	&	3.82	&	1.89	&	5.20	&	764	\\
\ref{fig:tube}	&	Glass	&	0.5	&	PMMA	&	20	&	n.a.	&	-2.37	&	-2.33	&	1.47	&	-0.34	&	0.72	&	556	\\
\ref{fig:tube}	&	Glass	&	0.5	&	PMMA	&	80	&	n.a.	&	-2.39	&	-1.68	&	1.96	&	-0.99	&	0.59	&	399	\\
\ref{fig:tube}	&	Glass	&	0.5	&	PMMA	&	100	&	n.a.	&	-3.55	&	-2.81	&	2.77	&	-1.38	&	1.96	&	525	\\
\ref{fig:RH}	&	Glass	&	0.5	&	Steel	&	40	&	20	&	-0.74	&	-0.56	&	0.58	&	-2.06	&	5.60	&	429	\\
\ref{fig:RH}, \ref{fig:matcomb}	&	Glass	&	0.5	&	Steel	&	40	&	30	&	-0.61	&	-0.56	&	0.29	&	-1.02	&	2.03	&	1474	\\
\ref{fig:RH}	&	Glass	&	0.5	&	Steel	&	40	&	60	&	-0.59	&	-0.44	&	0.46	&	-1.85	&	4.32	&	323	\\
\ref{fig:matcomb}	&	Glass	&	0.5	&	PMMA	&	40	&	20	&	-0.87	&	-0.76	&	0.59	&	-1.00	&	2.66	&	998	\\
\ref{fig:matcomb}	&	PS	&	0.5	&	PMMA	&	40	&	25	&	-0.36	&	-0.35	&	0.10	&	-0.86	&	2.34	&	406	\\
\ref{fig:matcomb}	&	PS	&	0.5	&	Steel	&	40	&	35	&	-0.49	&	-0.49	&	0.17	&	-0.07	&	3.59	&	686	\\
\ref{fig:matcomb}	&	Glass	&	0.5	&	PTFE	&	40	&	n.a.	&	0.52	&	0.43	&	0.48	&	0.46	&	4.69	&	134	\\
\hline\hline
\end{tabular}

\end{tabularx}
\end{table*}
\endgroup